\documentclass[a4paper,12pt]{article}
\usepackage{graphicx}
\global\arraycolsep=2pt 
\newlength{\orbarwd}\newlength{\orbarht}\newsavebox{\orbararg}%
\newcommand{\orbar}[1]{\savebox{\orbararg}{\ensuremath{#1}}%
  \settowidth{\orbarwd}{\usebox{\orbararg}}%
  \settoheight{\orbarht}{\usebox{\orbararg}}%
  \raisebox{1.1\orbarht}[0pt]{\makebox[0pt][l]{%
    \resizebox{1.1\orbarwd}{0.5ex}{\boldmath\ensuremath{(-)}}}}%
  \usebox{\orbararg}}%
\newcommand{\bra}[1]{\left\langle#1\right|}%
\newcommand{\ket}[1]{\left|#1\right\rangle}%
\newcommand{\bracket}[3]{\left\langle#1\left|#2\right|#3\right\rangle}%
\newlength{\eqnwidth}
\newenvironment{eqnacn}[1]{%
  \begin{equation}\label{#1}\begin{minipage}{\eqnwidth}%
    \renewcommand{\\}{\\\vspace{5mm}}%
    \vspace{-\abovedisplayskip}\begin{eqnarray*}}{%
  \end{eqnarray*}\end{minipage}\end{equation}}%
\newenvironment{eqnacn*}[1]{%
  \vspace{-\abovedisplayskip}\begin{eqnacn}{#1}}{\end{eqnacn}}%
\newcommand{\re}[1]{{\rm Re}\left\{#1\right\}}
\newcommand{\im}[1]{{\rm Im}\left\{#1\right\}}
\newcommand{\Br}[2]{\ensuremath{{\rm Br}\left(#1\to#2\right)}}
\newcommand{\abs}[1]{\left|#1\right|}%
\newcommand{\absq}[1]{\abs{#1}^2}%
\newcommand{\vma}[1]{\gamma#1(1\!-\!\gamma_5)}
\begin{document}
\makeatletter
\def\fmslash{\@ifnextchar[{\fmsl@sh}{\fmsl@sh[0mu]}}
\def\fmsl@sh[#1]#2{%
  \mathchoice
    {\@fmsl@sh\displaystyle{#1}{#2}}%
    {\@fmsl@sh\textstyle{#1}{#2}}%
    {\@fmsl@sh\scriptstyle{#1}{#2}}%
    {\@fmsl@sh\scriptscriptstyle{#1}{#2}}}
\def\@fmsl@sh#1#2#3{\m@th\ooalign{$\hfil#1\mkern#2/\hfil$\crcr$#1#3$}}
\makeatother
%
\thispagestyle{empty}
\begin{titlepage}

\begin{flushright}
Phys.\ Rev.\ {\bf D62}, 0960xx \\
hep-ph/0001022\\
TTP99--53\\
LMU 18/99\\
21 September 2000
\end{flushright}

\vspace{0.3cm}
\boldmath
\begin{center}
\Large\bf One-particle inclusive $CP$ asymmetries
\end{center}
\unboldmath
\vspace{0.8cm}

\begin{center}
{\large Xavier Calmet}\\
{\sl Ludwig-Maximilians-Universit\"at,
     Sektion Physik,\\
     Theresienstra\ss e 37, D--80333 M\"unchen, Germany}

\vspace{5mm}
{\large Thomas Mannel and Ingo Schwarze}\\
{\sl Institut f\"ur Theoretische Teilchenphysik,\\
     Universit\"at Karlsruhe, D--76128 Karlsruhe, Germany}

\vspace{\fill}
To be published in Physical Review {\bf D62}, 0960xx \\
(Received 5 January 2000; to be published 1 November 2000)
\end{center}

\vspace{\fill}
\begin{abstract}
\noindent
One-particle inclusive $CP$ asymmetries in the decays of the type $B
\to \orbar{D}^{(*)}X$ are considered in the framework of a QCD based
method to calculate the rates for one-particle inclusive decays.

\end{abstract}
\end{titlepage}

\setcounter{page}{2}
\thispagestyle{empty}
\mbox{}
\clearpage

\setcounter{page}{1}
\section{Introduction}
One of the main goals in $B$ physics is a detailed study of flavor
mixing, which is encoded in the Cabibbo-Kobayashi-Maskawa (CKM) matrix
of the standard model.  In particular, the violation of the $CP$
symmetry, which the standard model describes by a nontrivial phase in
the CKM matrix or equivalently by the angles of the unitarity
triangle, will be investigated.

Typically $CP$ asymmetries are expected to be large in some of the
exclusive nonleptonic $B$ decays which, however, have only small
branching ratios. Examples are the determination of $\beta$ from $B
\to J\!/\!\psi\,K_s$ and of $\alpha$ from $B \to \pi\pi$.  In
addition, in these exclusive nonleptonic decays it is very hard to
obtain a good theoretical control over the hadronic uncertainties, in
particular due to the presence of strong phases.

On the other hand, inclusive decays have large branching fractions but
typically smaller $CP$ asymmetries than exclusive decays
\cite{Dunietz99}.  One may use parton hadron duality to obtain a good
theoretical description. This has been studied by Beneke, Buchalla and
Dunietz who set up a theoretically clean method to calculate the $CP$
asymmetries in inclusive $B$ decays~\cite{BBD97}.  They still find
sizable $CP$ asymmetries, but their measurement would require to
identify charmless final states inclusively, which is not an easy
task.

One-particle inclusive decays lie somehow between these two cases.
This class of decays still has large branching fractions and some of
the expected $CP$ asymmetries are sizable. Furthermore, a measurement
of these decays is feasible.

For one-particle inclusive decays of the type $B \to
\orbar{D}^{(*)}X$, a QCD based description has been developed
recently, exploiting factorization and the heavy mass limit for both
the $b$ and the $c$ quark~\cite{CMS99}.  Since the expansion
parameters are $\Lambda_{\rm QCD}/(m_b-m_c)$, $1/N_C$ and
$\alpha_s(m_c)$, corrections to the leading term could be fairly
large, in the worst case of the order of $30\%$.  Using this method,
which unfortunately is not completely model independent, we compute
mixing induced time-dependent and time-integrated $CP$ asymmetries in
the framework of the standard model.

In view of the considerable uncertainties due to an unknown strong
phase, our method cannot yet be used for a competitive determination
of the $CP$ violation parameters, in particular compared to a
measurement of $\sin(2\beta)$ in the ``gold-plated'' channel $B \to
J\!/\!\psi\,K_s$.  However, it can be used as an estimate of the
one-particle inclusive $CP$ asymmetries, for which we shall use
present central values of the $CP$ angles $\beta$ and $\gamma$
\cite{Ciuchini99}.  Compared to fully inclusive methods, the advantage
is that we can predict asymmetries for the various spins and charges
of the ground-state charmed mesons separately.  This is certainly a
worthwhile task, in particular since we are not aware of any previous
prediction for these asymmetries, not even in the context of quark
models.

After introducing our notations for $B$ mixing in Sec.~\ref{sec:asym},
we calculate the relevant matrix elements in Sec.~\ref{sec:matrix} and
model the form factors in Sec.~\ref{sec:formfact}.  The numerical
results are given in Sec.~\ref{sec:results}.

\section{\boldmath $CP$ asymmetries in
         $B \to \protect\orbar{D}^{(*)}X$} \label{sec:asym}
In Wigner Weisskopf approximation the time evolution of an initially
pure $B^0$ or $\overline B^0$,
\begin{eqnacn*}{eq:bsystem}
  \ket{B^0_{\rm phys}(t)} & = &
    g_+(t) \ket{B^0} - \frac{q}{p} g_-(t) \ket{\overline B^0}, \\
  \ket{\overline B^0_{\rm phys}(t)} & = &
    g_+(t) \ket{\overline B^0} - \frac{p}{q} g_-(t) \ket{B^0},
\end{eqnacn*}%
is determined by the time-dependent functions
\begin{eqnacn}{eq:gfunc}
  g_+(t) & = & e^{-iMt -\frac{1}{2} \Gamma t}
    \left[\cosh\frac{\Delta\Gamma t}{4} \cos\frac{\Delta Mt}{2}
      + i \sinh\frac{\Delta{\Gamma}t}{4} \sin\frac{\Delta Mt}{2}
    \right] \\
  g_-(t) & = & e^{-iMt -\frac{1}{2} \Gamma t}
    \left[\sinh\frac{\Delta\Gamma t}{4} \cos\frac{\Delta Mt}{2}
      + i \cosh\frac{\Delta\Gamma t}{4} \sin\frac{\Delta Mt}{2}
    \right],
\end{eqnacn}%
where $\Delta M = M_H - M_L > 0$ and $\Delta\Gamma = \Gamma_H -
\Gamma_L < 0$ are the mass and width differences between the mass
eigenstates $|B_H\!\!> \, = p |B^0\!\!> + \: q |\overline B^0\!\!>$ and
$|B_L\!\!> \, = p |B^0\!\!> - \: q |\overline B^0\!\!>$.

The quantity $q/p$ is given in terms of the off-diagonal elements of
the Hamiltonian $H = M - i\Gamma/2$  of the neutral $B$ meson
system
\begin{equation}
  \frac{q}{p} =
  \frac{\Delta M - \frac{i}{2}\Delta\Gamma}
       {2\left(M_{12} - \frac{i}{2}\Gamma_{12}\right)} =
  \frac{M_{12}^*}{|M_{12}|}
    \left( 1 - \frac{1}{2} a + {\cal O}(a^2) \right), \qquad
  a = \im{ \frac{\Gamma_{12}}{M_{12}} }.
\end{equation}
In fact, $\Gamma_{12}/M_{12}={\cal O}(m_b^2/m_t^2)$ is very small and
hence $q/p$ is to a good approximation a phase factor.

The time-dependent rate for the decay of a $B$ meson into a set of
final states $\ket{f} = \sum_i\ket{f_i}$ can be written as
\begin{eqnarray}
  \nonumber \Gamma [B(t) \! \to \! f] & = &
  \frac{1}{2m_B} \sum_i \int\!d\phi_i\,(2\pi)^4 \delta^4 (p_B-p_{f_i})
    \bracket{B(t)}{H_{\rm eff}}{f_i}\bracket{f_i}{H_{\rm eff}}{B(t)} \\
  & = & \frac{1}{2m_B} \int d^4 x \,
    \bracket{B(t)}{H_{\rm eff}(x) \Pi_f H_{\rm eff}(0)}{B(t)},
\end{eqnarray}
where $d\phi_i$ is the phase space element of the state $\ket{f_i}$
and
\begin{equation}
  \Pi_f = \sum_i \int d\phi_i \, \ket{f_i}\bra{f_i}
\end{equation}
is the projector on the set of final states.  Note that both an
exclusive final state as well as inclusive states can be treated in
this way.  Even differential distributions can be considered if the
phase spaces $d\phi_i$ are not fully integrated.

The $CP$ asymmetries we are going to consider are of the type
\begin{equation}
  {\cal A}_{CP}(t) =
    \frac{\Gamma(B^0(t)\to f) - \Gamma(\overline B^0(t)\to\overline f)}
         {\Gamma(B^0(t)\to f) + \Gamma(\overline B^0(t)\to\overline f)}
\end{equation}
which involves the $CP$ conjugate set $|\overline f\!>$ of final
states.

Up to here the discussion is completely general.  In the following we
shall use the above formalism to compute the $CP$ asymmetries for
one-particle inclusive final states, for which the projector reads
\begin{equation}
  \Pi_f = \sum_X \ket{XY}\bra{XY},
\end{equation}
where $Y$ can be a $D$ or a $\overline D$ meson.  Since the sum runs
over all possible states $X$, the $CP$ conjugate of the projector is
\begin{equation}
  \Pi_{\overline f} = \sum_X \ket{X\overline Y}\bra{X\overline Y}.
\end{equation}
Inserting the time-dependent states~(\ref{eq:bsystem}) we obtain
\begin{eqnacn}{eq:rates}
  \!\!\! \Gamma [B(t) \!\! \to \!\! YX] & = &
    \absq{g_+(t)} \Gamma_Y^{BB} +
    \absq{\frac{q}{p} g_-(t)} \! \Gamma_Y^{\overline B \, \overline B} -
    2 \re{\frac{q}{p} g_+^*g_-(t) T_Y^{B \overline B} }, \!\!\! \\
  \!\!\! \Gamma [\overline B(t) \!\! \to \!\! \overline YX] & = &
    \absq{g_+(t)} \Gamma_{\overline Y}^{\overline B \, \overline B} +
    \absq{\frac{p}{q} g _-(t)} \! \Gamma_{\overline Y}^{BB} -
    2 \re{\frac{p}{q} g_+^*g_-(t) T_{\overline Y}^{\overline BB}},
    \!\!\!
\end{eqnacn}%
where the matrix elements are defined by
\begin{eqnacn}{eq:tdef}
  \Gamma_{Y}^{BB} & = & \frac{1}{2m_B} \int d^4 x \,
    \bracket{B}{H_{\rm eff}(x) \Pi_Y H_{\rm eff}(0)}{B}, \\
  T_{Y}^{B \overline B} & = & \frac{1}{2m_B} \int d^4 x \,
    \bracket{B}{H_{\rm eff}(x) \Pi_Y H_{\rm eff}(0)}{\overline B}.
\end{eqnacn}%
The $\Delta B = 2$ transition matrix elements representing the
interference between the mixed and the unmixed amplitudes are related
by $CPT$ symmetry, such that
\begin{equation}
  T_Y := T_Y^{B \overline B} = \left( T_Y^{\overline BB} \right)^*.
\end{equation}

The direct $CP$ asymmetries in these processes are expected to be
tiny.  In fact, using the method described in Ref.~\cite{CMS99}, they
turn out to be of higher order in the $1/m$ expansion.  Hence we have
\begin{eqnarray} \label{eq:gsym}
  \Gamma_Y  :=   \Gamma_Y^{BB}
           & = & \Gamma_{\overline Y}^{\overline B \, \overline B}
             =   \Gamma(B \to YX), \qquad 
  \Gamma_{\overline Y} := \Gamma_{\overline Y}^{BB}, \\[2mm]
       T_Y & = &      T_{\overline Y}.
\end{eqnarray}

Inserting the time-dependent decay rates in Eq.~(\ref{eq:rates}) and
neglecting both the width difference and $a$, such that $q/p$ becomes
a phase factor, we obtain for the time-dependent $CP$ asymmetries
\begin{equation} \label{eq:atime}
  {\cal A}_{CP}(t) = \frac
  { \sin\left( \Delta Mt \right) \im{ \frac{q}{p} T_Y } }
  { \cos^2 \left( \frac{\Delta Mt}{2} \right) \Gamma_Y +
    \sin^2 \left( \frac{\Delta Mt}{2} \right) \Gamma_{\overline Y} },
\end{equation}
from which we get the time-integrated asymmetry
\begin{equation} \label{eq:aint}
  {\cal A}_{CP} = \frac
  { 2 \, x \, \im{ \frac{q}{p} T_Y } }
  { \left( 2 + x^2 \right) \Gamma_Y + x^2 \, \Gamma_{\overline Y} },
\end{equation}
where $x = \Delta M/\Gamma$ is measured to be $x = 0.73$ \cite{PDG}.

\section{Transition matrix elements} \label{sec:matrix}
In order to compute the $CP$ asymmetries, one has to evaluate the
matrix elements in Eq.~(\ref{eq:tdef}).  The total rates $\Gamma_Y$
have already been discussed in Ref.~\cite{CMS99}, so we only need to
calculate the interference term $T_Y$.

The relevant pieces of the effective Hamiltonian contributing to this
interference are $(\overline ub)_{V-A} (\overline dc)_{V-A}$ and
$(\overline cb)_{V-A} (\overline du)_{V-A}$ interfering with each
other and $(\overline cb)_{V-A} (\overline dc)_{V-A}$ interfering with
itself, so $T_Y$ is a sum of the two contributions
\begin{eqnarray}
  T_Y & = & T_c + T_u, \\
  T_q & = & \frac{1}{2m_B} \frac{G_F^2}{2}
    V^{ }_{cb} V^*_{qd} V^{ }_{qb} V^*_{cd}
    \absq{C_1} \sum_X (2\pi)^4 \delta^4(p_B-p_D-p_X)
  \\ \nonumber & &
  \bracket{B^0}{(\overline qb)_{V-A}(\overline dc)_{V-A}}{DX}
  \bracket{DX}{(\overline dq)_{V-A}(\overline cb)_{V-A}}{\overline B^0}.
\end{eqnarray}

Fierzing the operators into the form $(\overline db)_{V-A} (\overline
uc)_{V-A}$, $(\overline db)_{V-A} (\overline cu)_{V-A}$ and
$(\overline db)_{V-A} (\overline cc)_{V-A}$ one can reproduce the
inclusive results of Ref.~\cite{BBD97}.  In order to evaluate the
interference term for the one-particle inclusive case, we use the
method developed in Ref.~\cite{CMS99}.  It is based on factorization,
which holds to leading order in the $1/N_C$ expansion, where $N_C$ is
the number of QCD colors.  Thus we can write the interference terms as
products of two tensors
\begin{equation}
  T_q = \frac{1}{2m_B} \frac{G_F^2}{2}
    V^{ }_{cb} V^*_{qd} V^{ }_{qb} V^*_{cd}
    \absq{C_1} \int \! \frac{d^4 Q}{(2\pi)^4} \,
    K_{\mu\nu}(p_B,Q) \int \! d\phi_D \, P_q^{\mu\nu}(p_D,Q)
\end{equation}
with
\begin{eqnarray}
  K_{\mu\nu}(p_B,Q) & = & \sum_X (2\pi)^4 \delta^4(p_B-p_X-Q)
  \\ \nonumber &&
  \bracket{B^0(p_B)}{(\overline d \vma{_\mu} b)}{X}
  \bracket{X}{(\overline d \vma{_\nu} b)}{\overline B^0(p_B)},
  \\[2mm]
  P_q^{\mu\nu}(p_D,Q) & = & \sum_{X'} (2\pi)^4 \delta^4(Q-p_D-p_{X'})
  \\ \nonumber &&
  \bracket{0}{(\overline q \vma{^\mu} c)}{D^{(\!*\!)}(p_D) X'}
  \bracket{D^{(\!*\!)}(p_D) X'}{(\overline c \vma{^\nu} q)}{0}.
\end{eqnarray}

The tensor $K_{\mu\nu}(p_B,Q)$ is fully inclusive and one can perform
a standard short distance expansion.  The resulting $\Delta B=2$
matrix element can be parameterized by the decay constant $f_B$ of the
$B$ meson and the bag factors $B$ and $B_s$ for the axial vector and
the scalar current, respectively.

The other tensor $P_q^{\mu\nu}(p_D,Q)$ involves a projection on a
one-particle inclusive charmed meson state and hence we cannot perform
a short distance expansion.  We proceed along the same lines as in
Ref.~\cite{CMS99}, where the rates for wrong charm decays have been
modeled.  Heavy quark symmetry yields the Dirac matrix structure
\begin{equation}\label{eq:hqs}
  P_q^{\mu\nu}(p_D,Q) \propto \overline H_{D^{(*)}}(p_D) \vma{^\mu}
  \otimes \vma{^\nu} H_{D^{(*)}}(p_D),
\end{equation}
where the representation matrices for the charmed mesons are
\begin{equation}
  H_D = \sqrt{m_D} \, \frac{1+\fmslash{v}_D}{2} \, \gamma_5, \qquad
  H_{D^*} = \sqrt{m_{D^*}} \, \frac{1+\fmslash{v}_{D^*}}{2} \,
            \fmslash{\epsilon}.
\end{equation}
In principle, all possible contractions of the light quark indices may
contribute, giving rise to several form factors.  For a first
estimate, it is sufficient to use only the simplest one of these
contractions,
\begin{equation}\label{eq:contr}
  P_q^{\!\mu\nu}(p_D,\!Q) \!=\! 2\pi \,
    \delta \! \left( \left( Q \!-\! p_D \right)^2 \!\!-\! m_q^2 \right)
    {\rm Tr} \Big\{ \fmslash{p}_D \, \vma{^\mu}
      \left( \fmslash{Q} \!-\! \fmslash{p}_D \right) \vma{^\nu} \!\Big\}
    \tilde f_{qY},
\end{equation}
corresponding to a replacement of the $D^{(*)}X$ final state by a pair
of free quarks, rescaled by an operator- and decay-channel-specific
form factor $\tilde f_{qY}$, where $Y$ is one of the ground state $D$
mesons.  In the following, we call this contraction ``partonic.''

Using this ansatz and the heavy mass limit, the transition matrix
elements read
\begin{eqnarray}
\label{eq:tc}
T_c & = & - \frac{G_F^2 m_B^3 f_B^2}{24 \pi}
          (V^{ }_{cb} V^*_{cd})^2 \absq{C_1} \sqrt{1\!-\!4z}
          \big[ (1\!-\!4z)B \!+\! 2(1\!+\!2z)B_S \big] \tilde f_{cY}, \\
\label{eq:tu}
T_u & = & - \frac{G_F^2 m_B^3 f_B^2}{24 \pi}
          V^{ }_{cb} V^*_{ud} V^{ }_{ub} V^*_{cd}
          \absq{C_1} (1\!-\!z)^2
          \big[ (1\!-\!z) B \!+\! 2 (1\!+\!2z) B_S \big]
          \tilde f_{uY}, \quad
\end{eqnarray}
where $z=(m_c/m_b)^2$ and $C_1$ is the Wilson coefficient of the
effective Hamiltonian in the notation of Ref.~\cite{CMS99}.
Equations~(\ref{eq:tc}) and (\ref{eq:tu}) correspond to the expression
for the width difference of neutral heavy meson
systems~\cite{Hagelin81}.

In the standard CKM parametrization, the phases of the transition
matrix elements are
\begin{eqnarray}
  \arg(T_c) & = & 0, \\
  \arg(T_u) & = & \arg( - V^{ }_{ub} ) = - \gamma, \\[2mm]
  \arg(q/p) & = & \arg( - V^2_{td} ) = - 2 \beta,
\end{eqnarray}
such that
\begin{equation}
  \im{ \frac{q}{p} T_Y } = \sin(2\beta)\abs{T_c}
                         + \sin(2\beta+\gamma)\abs{T_u}.
\end{equation}

\section{Modeling the form factors} \label{sec:formfact}
We assume that the form factors $\tilde f_{qY}$ do not vary strongly
over the accessible phase space and hence we approximate them by
constants.  For the case $q=c$, these constants have been fitted to
the wrong charm yield in $B$ decays~\cite{CMS99}.  Operators analogous
to the case $q=u$ are Cabibbo suppressed when calculating wrong charm
rates, so they did not appear in Ref.~\cite{CMS99}.  Assuming that all
charm quarks eventually hadronize to $D$ mesons, we use
\begin{equation} \label{eq:fsat}
  \tilde f_{uD^0} + \tilde f_{uD^+} = 1.
\end{equation}

To resolve the spin and charge counting, we first discuss the heavy
mass limit where the pseudoscalar and vector charmed mesons form a
degenerate ground state doublet.  The decay of vector to pseudoscalar
mesons will be discussed below.  In the following, $D_{\rm dir}$
refers to those $D$ mesons that do not result from $D^*$ decays, and
$D^{(*)}$ can be either $D_{\rm dir}$ or $D^*$.

As long as the light quark spin indices of the $D^{(*)}$ meson
representation matrices are contracted with each other,
Eq.~(\ref{eq:hqs}) reproduces the naive spin counting
\begin{equation}\label{eq:spin}
  \tilde f_{qD^{*0}} = 3 \tilde f_{qD^0_{\rm dir}}, \qquad
  \tilde f_{qD^{*+}} = 3 \tilde f_{qD^+_{\rm dir}}.
\end{equation}
Different contractions yield results of comparable size.
The experimental spin counting factor appears to be smaller by roughly
a factor of two~\cite{CMS99}.  Since this effect is not yet
understood, we treat it as an uncertainty.

\begin{figure}
  \begin{center}
    \includegraphics[scale=0.5]{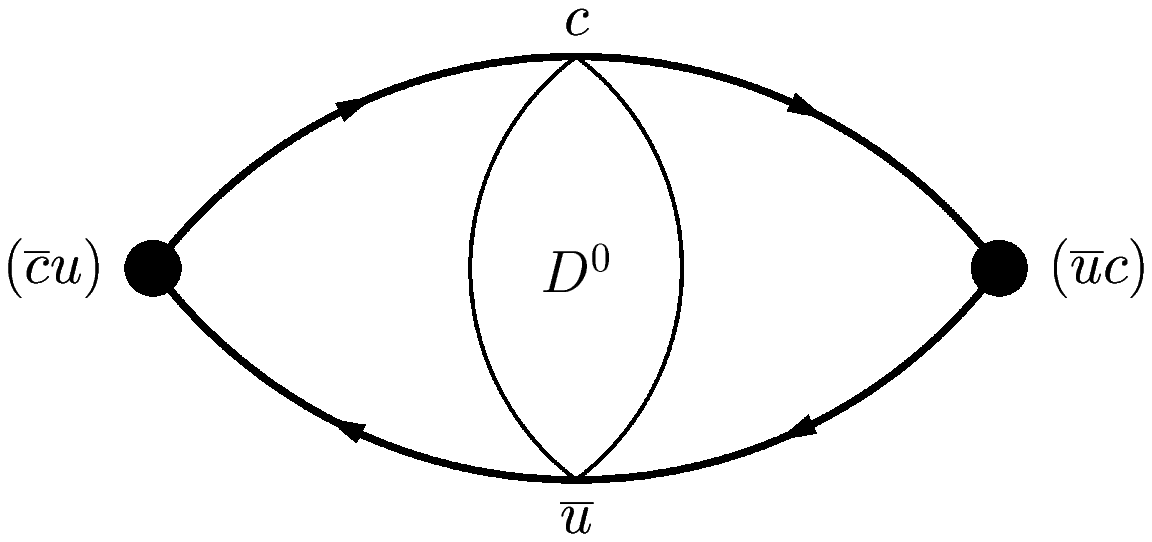}
    \caption{Topology yielding $\absq{a_1}$
             in Eq.~(\ref{eq:interfere}).
             \label{fig:spectator}}
  \end{center}
  \begin{minipage}[t]{0.49\linewidth}\centering
    \includegraphics[scale=0.5]{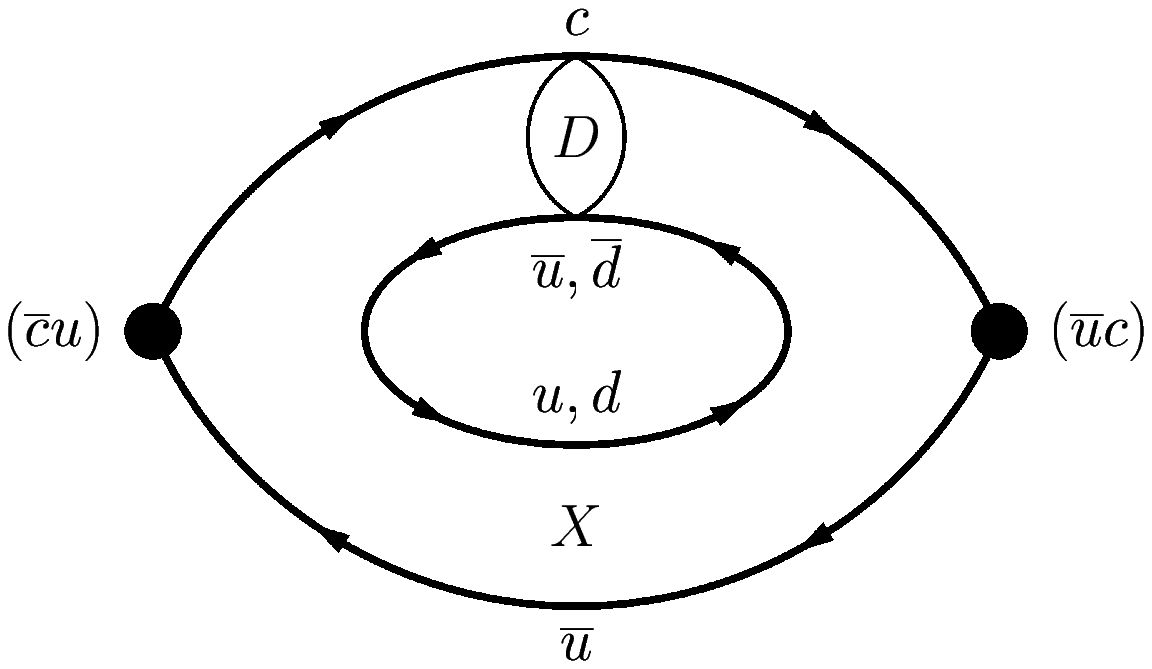}
    \begin{minipage}{0.9\linewidth}
      \caption{Topology yielding $\absq{a_2}$
               in Eq.~(\ref{eq:interfere}).
               \label{fig:pop}}
    \end{minipage}
  \end{minipage}
  \begin{minipage}[t]{0.49\linewidth}\centering
    \includegraphics[scale=0.5]{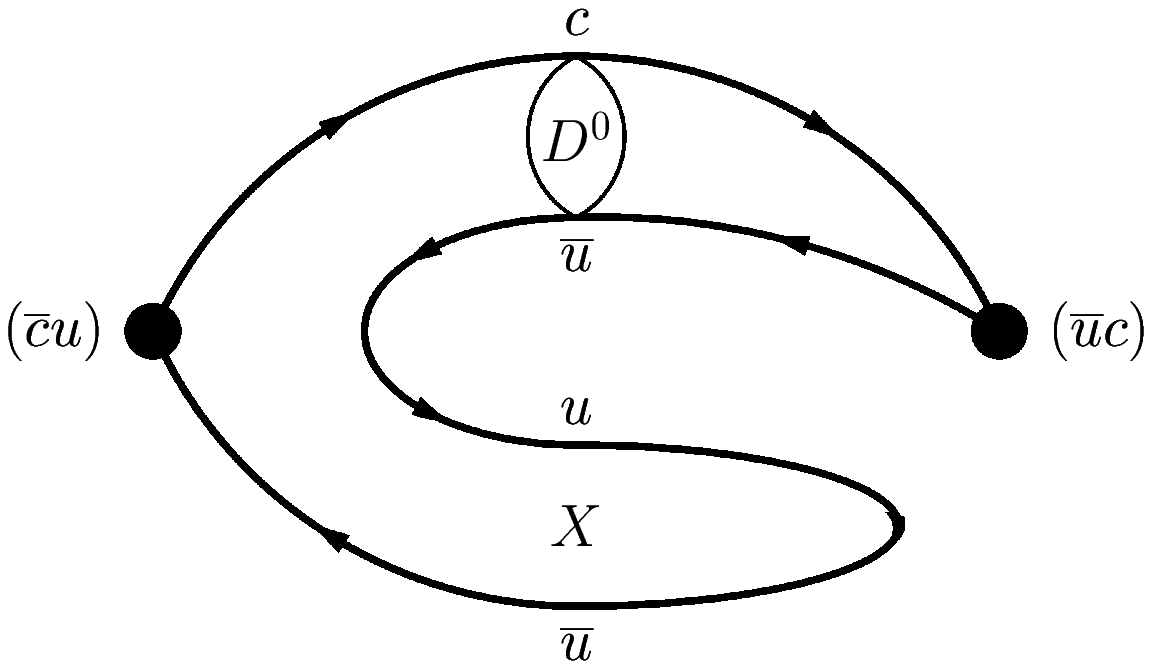}
    \begin{minipage}{0.9\linewidth}
      \caption{Interference topology for Eq.~(\ref{eq:interfere}).
               \label{fig:interfere}}
    \end{minipage}
  \end{minipage}
\end{figure}

Concerning charge counting, we argued by isospin symmetry~\cite{CMS99}
that in the case $q=c$ we have
\begin{equation}
  \tilde f_{cD^{(*)0}} = \tilde f_{cD^{(*)+}}.
\end{equation}
In the case $q=u$, two topologies can contribute to the decay
amplitude: the charm quark can either hadronize with the $u$ quark
from the weak effective current, in which case the isospin of the
state $\ket{X}$ is $I_X = 0$, or with a $u$ or $d$ quark from vacuum,
which contains both $I_X = 0$ and $I_X = 1$ contributions.  In the
case $I_X = 0$, both amplitudes can interfere, so there are three
contributions to the decay rate
\begin{eqnacn}{eq:interfere}
  \tilde f_{uD^{(*)0}} & = & \absq{a_1+a_2} =
    \absq{a_1} + \absq{a_2} + 2 \re{a_1^* a_2} \\
  \tilde f_{uD^{(*)+}} & = & \absq{a_2},
\end{eqnacn}
see Figs.~\ref{fig:spectator}--\ref{fig:interfere}.

One might doubt whether using the partonic contraction given in
Eq.~(\ref{eq:contr}) is justified for all the topologies, as it
appears to correspond to the topology in Fig.~\ref{fig:pop}, while the
topology in Fig.~\ref{fig:spectator} should rather be described by the
contraction
\begin{equation}
  P_q^{\mu\nu}(p_D,Q) \propto
  {\rm Tr} \Big\{ \overline H_{D^{(*)}}(p_D) \vma{^\mu} \Big\}
  {\rm Tr} \Big\{ \vma{^\nu} H_{D^{(*)}}(p_D) \Big\}.
\end{equation}

This is not a problem for three reasons.  First, we do not claim to be
able to accurately model the matrix element, but we only give the
simplest possible ansatz by rescaling the partonic result.  In
particular, it is clearly not yet feasible to model particular
contributions individually.  We only use the three topologies to
estimate the integrated relative magnitudes of the two main
contributions and to bound the magnitude of their interference term.
Secondly, neither the time-dependent nor the time-integrated
asymmetries depend on the choice of the contraction unless studied
differentially in the momentum of the charmed meson, which so far we
do not attempt to do.  Finally, as noted in Ref.~\cite{CMS99}, the
choice of the wrong charm contraction appeared to have little
influence even on differential observables.

The topologies in Figs.~\ref{fig:spectator} and \ref{fig:pop} also
occur in wrong charm production in $B$ decays.
Figure~\ref{fig:spectator} corresponds to the process $B \to
D_s^{(*)+}X$, Fig.~\ref{fig:pop} to the process $B \to D^{(*)}X$,
where $D^{(*)}$ can be either $D^{(*)0}$ or $D^{(*)+}$.  Both
contributions are experimentally known to be of similar size, i.e.,
$(10 \pm 2.5)\%$ \cite{PDG} and $(7.9 \pm 2.2)\%$ \cite{Coan98},
respectively, such that
\begin{equation}
  \absq{a_1} = 2 \absq{a_2}.
\end{equation}
The relative phase of the two contributions is unknown.  Therefore,
although it may be large, we have to treat the interference part as a
theoretical uncertainty.  This is acceptable since the $q=u$
contribution is smaller than the $q=c$ contribution according to
\begin{equation}
  \abs{\frac{T_u}{T_c}} =
  \abs{\frac{V_{ub}}{V_{cb}}\frac{V_{ud}}{V_{cd}}}
  \frac{(1-z)^2 (1+z)}{\sqrt{1-4z}} \frac{\tilde{f}_u}{\tilde{f}_c}
  \propto \abs{\frac{V_{ub}}{V_{cb}}} \frac{(1+z)}{\abs{V_{cd}}}
  \approx 0.4.
\end{equation}

Off the heavy mass limit, $D^* \to D$ decay has to be taken into
account.  In the same way as in Ref.~\cite{CMS99}, we get
\begin{eqnacn}{eq:feeddown}
  \tilde f_{qD^+} & = &
    \tilde f_{qD^+_{\rm dir}} +
    \Br{D^{*+}}{D^+ X} \tilde f_{qD^{*+}} \\
  \tilde f_{qD^0} & = &
    \tilde f_{qD^0_{\rm dir}} + \tilde f_{qD^{*0}} +
    \Br{D^{*+}}{D^0 X} \tilde f_{qD^{*+}}.
\end{eqnacn}%
The coefficients obtained from
Eqs.~(\ref{eq:fsat})--(\ref{eq:feeddown}) and Ref.~\cite{CMS99} are
summarized in Table~\ref{tab:formfact}.  The ranges given result from
varying the spin counting factor in Eq.~(\ref{eq:spin}) from $3$ down
to $3/2$ and the interference in Eq.~(\ref{eq:interfere}) from the
central value of vanishing interference to full constructive and
destructive interference.

\begin{table}
\centering
\newlength{\ld}\settodepth{\ld}{g}\addtolength{\ld}{3pt}
\settoheight{\unitlength}{l}
  \addtolength{\unitlength}{\ld}\addtolength{\unitlength}{4pt}
\newcommand{\mh}{\rule[-\ld]{0pt}{\unitlength}}
\renewcommand{\arraystretch}{0}
\begin{tabular}{|@{}c@{}|c|c|c|r@{--}l|}
  \hline
  \mh & \multicolumn{2}{|c|}{$q=c$} & \multicolumn{3}{|c|}{$q=u$} \\
  \raisebox{-\ld}[0pt][0pt]{
    \begin{picture}(4,2)
      \put(4,1.8){\makebox(0,0)[tr]{operator}}
      \put(0,0.2){channel}
      \put(0,2){\line(2,-1){4}}
    \end{picture}}
  \mh & central & r. to & central & \multicolumn{2}{|c|}{range} \\
  \hline
  \mh$\tilde f_{qD^+_{\rm dir}}$&$\:\:2/16$&$0.2$&$\:\:1/16$&0.04&0.34\\
  \mh$\tilde f_{qD^0_{\rm dir}}$&$\:\:2/16$&$0.2$&$\:\:3/16$&0.34&0.04\\
  \mh$\tilde f_{qD^{*+}}$       &$\:\:6/16$&$0.3$&$\:\:3/16$&0.09&0.64\\
  \mh$\tilde f_{qD^{*0}}$       &$\:\:6/16$&$0.3$&$\:\:9/16$&0.64&0.09\\
  \mh$\tilde f_{qD^+}$          &$\:\:4/16$&$0.3$&$\:\:2/16$&0.07&0.51\\
  \mh$\tilde f_{qD^0}$          &   $12/16$&$0.7$&   $14/16$&0.93&0.49\\
  \hline
\end{tabular}
\caption{Operator- and channel-specific form factors.
         \label{tab:formfact}}
\end{table}

\section{Results} \label{sec:results}
\begin{figure}\centering
  \includegraphics[width=\linewidth]{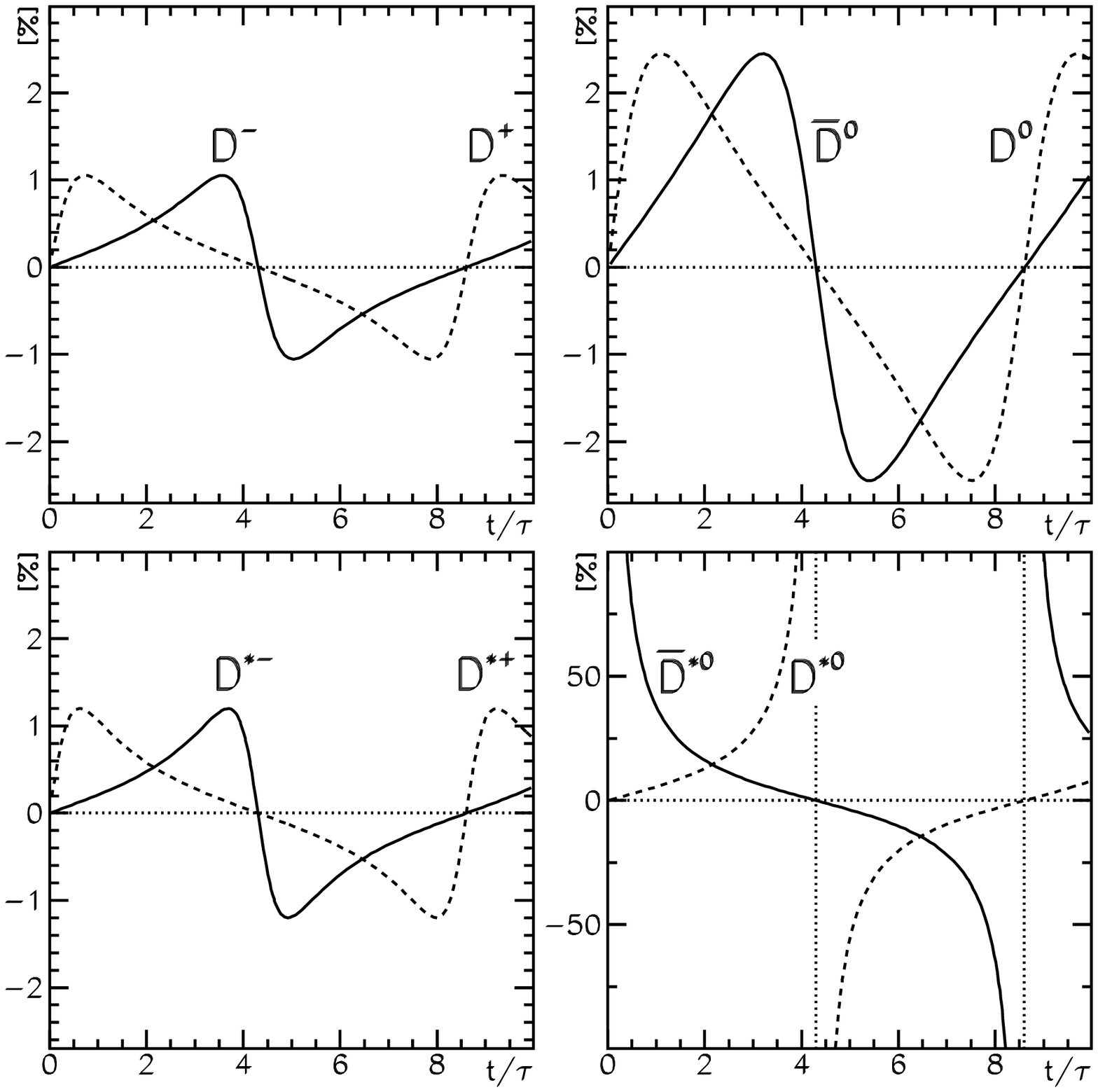}
  \caption{Time-dependent $CP$ asymmetries
    in $B^0 \to \protect\orbar{D}X$
    for pseudoscalar (above), vector (below),
    charged (left), neutral (right),
    right charm (solid), and wrong charm (dashed)
    $\protect\orbar{D}$ mesons.
    \label{fig:atime}}
\end{figure}

We have computed the parameters for the time-dependent $CP$
asymmetries as well as the time-integrated asymmetries.  We have
inserted recent values for $\sin 2\beta=0.75$ and $\gamma=68^\circ$
\cite{Ciuchini99}.  In addition, we use $V_{cb}=0.04$,
$V_{ub}=0.08\,V_{cb}$, $z=0.09$, $x=0.73$, $f_B=180$ MeV,
$\Br{D^{*+}}{D^0Y} = 1 - \Br{D^{*+}}{D^+ Y} = 0.683$ and
$C_1=B=B_S=1$.  The results of the calculations can be found in
Fig.~\ref{fig:atime} and Table~\ref{tab:aint}.

To assess the uncertainties involved in Fig.~\ref{fig:atime}, note
that according to Eq.~(\ref{eq:atime}) the shapes of the
time-dependent asymmetries are determined by the ratios of the wrong
to right charm rates $\Gamma_{\overline Y}/\Gamma_Y$.  We
checked numerically that the shapes would hardly change even if these
ratios were off by $30\%$.  The dominant contribution to the
uncertainty of the amplitudes arises from the transition matrix
elements $T_Y$ and is directly proportional to the uncertainties of
the time-integrated asymmetries given in Table~\ref{tab:aint}.

Suppose $N$ perfectly tagged $B^0$ decays are recorded in an
experiment.  In order to establish the asymmetry in a channel with a
branching ratio $b$ on the $3\sigma$ level,
\begin{equation} \label{eq:minumber}
  \frac{\cal A}{3} \ge \Delta {\cal A} = \frac{1}{\sqrt{2bN}}
\end{equation}
has to be satisfied.  The necessary numbers of tagged $B^0$ decays are
given in the last column of Table~\ref{tab:aint}.  Since the asymmetry
tends to be roughly inversely proportional to the branching ratio by
Eq.~(\ref{eq:aint}), we obtain from Eq.~(\ref{eq:minumber})
\begin{equation}
  N \propto \frac{1}{{\cal A}^2 b} \propto b,
\end{equation}
such that rare channels are advantageous for observing one-particle
inclusive asymmetries.

\begin{table}
\centering
\begin{tabular}{|l|r|c|r@{--}l|r|}
\hline
decay & Br \cite{CMS99} & ${\cal A}$ &
        \multicolumn{2}{|c|}{${\cal A}$ range} & necessary \\
channel & (\%) & (\%) & \multicolumn{2}{|c|}{(\%)} & $B^0$ decays \\
\hline
$B^0 \to D^- X$              & 29.1& 0.16& 0.15& 0.29& 6.000.000\\
$B^0 \to \overline D^0 X$    & 31.8& 0.58& 0.59& 0.46&   400.000\\
$B^0 \to D^+ X$              &  2.2& 0.58& 0.54& 1.04& 6.000.000\\
$B^0 \to D^0 X$              &  5.7& 1.53& 1.56& 1.23&   350.000\\
$B^0 \to D^{*-} X$           & 46.8& 0.16& 0.12& 0.23& 4.000.000\\[-2pt]
$B^0 \to \overline D^{*0} X$ &  (0)& (20)& (21)& (10)&$>$ 80.000\\
$B^0 \to D^{*+} X$           &  2.5& 0.61& 0.45& 0.89& 5.000.000\\
$B^0 \to D^{*0} X$           &  2.5& 4.17& 4.40& 2.20&   100.000\\
\hline
\end{tabular}
\caption{Branching ratios, integrated $CP$ asymmetries
         and numbers of necessary tagged $B^0$ decays
         for the one-particle inclusive
         $B^0 \to \protect\orbar{D}^{(*)}X$ decay channels.
         Concerning $\overline D^{*0}$, see the text.
         \label{tab:aint}}
\end{table}

The channel $B^0 \to \overline D^{*0}X$ deserves a further comment.
Looking at Fig.~\ref{fig:atime}, there is an obvious problem at small
proper decay times.  The reason for this problem is that we have
discussed all the rates only to leading order in the combined $1/N_C$
and $1/m_Q$ expansions.  However, this leading term vanishes for the
channel $B^0 \to \overline D^{*0}X$ and thus subleading terms become
relevant.  On the other hand, the numerator $T_Y$ of the $CP$
asymmetries is given by a matrix element of a dimension six operator
and hence is suppressed compared to the leading terms of most of the
rates. In other words, while in most of the rates the asymmetries are
of subleading order $f_B^2 / m_B^2$, this is not the case for the
channel $B^0 \to \overline D^{*0}X$.

\begin{figure}
  \begin{minipage}[t]{0.49\linewidth}\centering
    \includegraphics[width=\linewidth]{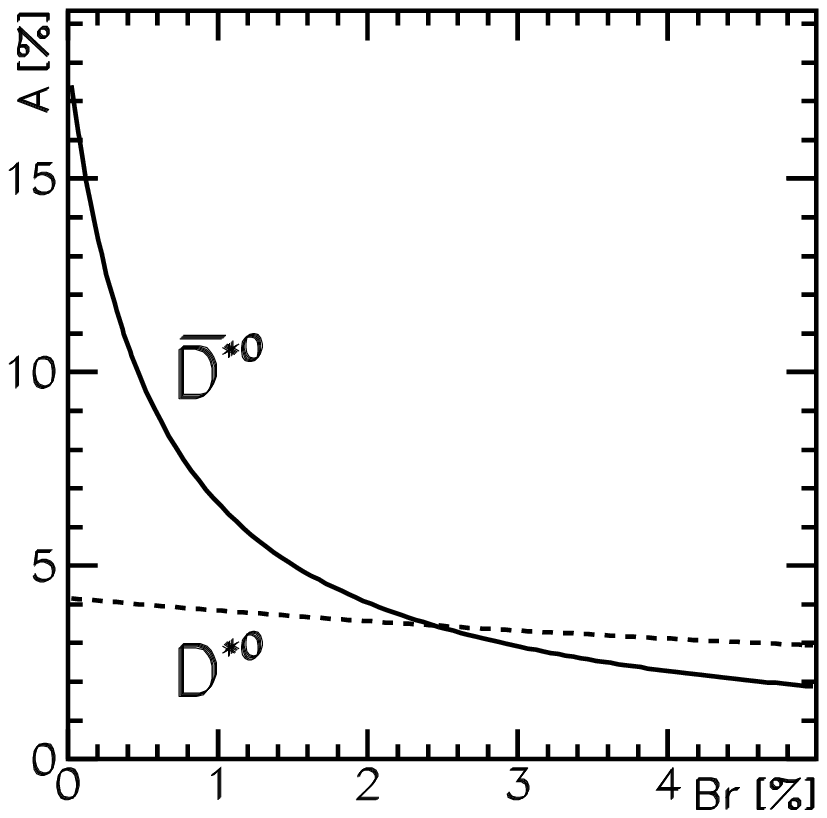}
    \begin{minipage}{0.9\linewidth}
      \caption{Time-integrated asymmetry in
        $B^0 \to \protect\orbar{D}^{*0}X$
        as a function of ${\rm Br} (B^0 \to \overline D^{*0}X)$.
        \label{fig:ds0asym}}
    \end{minipage}
  \end{minipage}
  \begin{minipage}[t]{0.49\linewidth}\centering
    \includegraphics[width=\linewidth]{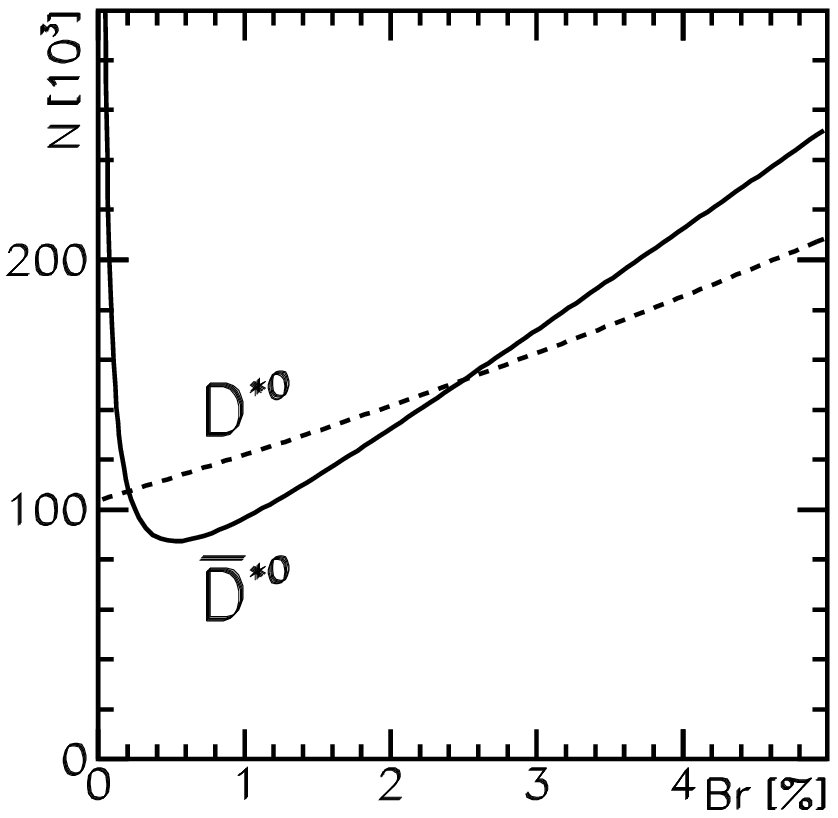}
    \begin{minipage}{0.9\linewidth}
      \caption{Necessary number of tagged $B^0$ events in
        $B^0 \to \protect\orbar{D}^{*0}X$
        as a function of ${\rm Br} (B^0 \to \overline D^{*0}X)$.
        \label{fig:ds0numb}}
    \end{minipage}
  \end{minipage}
\end{figure}

Unfortunately we cannot compute this possibly large asymmetry, since
this would involve to compute subleading terms for the decay rate.
Hence we try to estimate the asymmetry by varying $\Br{B^0}{\overline
D^{*0} X}$ in Eq.~(\ref{eq:aint}) and show the reaction of the
asymmetry in Fig.~\ref{fig:ds0asym} and of the necessary number of
tagged $B^0$ events in Fig.~\ref{fig:ds0numb}.  The wrong charm
asymmetry is practically unaffected by $\Br{B^0}{\overline D^{*0} X}$
since the pole occurs near four average lifetimes where most of the
$B$ mesons have already decayed, but the right charm asymmetry turns
out to be extremely sensitive.  Therefore we cannot predict the latter
quantitatively, but it can be as large as several percent, and it will
be measurable with a few $100\,000$ tagged $B^0$ events.

\section{Conclusion}
Motivated by the work on fully inclusive $CP$ asymmetries and the
question how to measure them, we studied one-particle inclusive $CP$
asymmetries.  In the final state only a $\orbar{D}^{(*)}$ meson has to
be identified and thus they are experimentally more easily accessible
than the fully inclusive $CP$ asymmetries.

We have used a similar method as in in Ref.~\cite{CMS99} to calculate
the time-dependent and time-integrated $CP$ asymmetries for
one-particle inclusive $B \to \orbar{D}^{(*)}X$ decays.  It turns out
that, as in Ref.~\cite{CMS99}, one cannot avoid to introduce some
model dependence.  Furthermore, there is also some dependence on an
unknown relative phase, which we treat as an uncertainty.  Due to
these uncertainties we cannot expect our method to compete with
proposed methods using ``gold-plated'' channels for determining CKM
parameters, but we can still give estimates for the expected $CP$
asymmetries of the different ground state $\orbar{D}$ mesons.

For most of the asymmetries we find results of a few $10^{-3}$, but
some are expected to be as large as several percent. These effects
should be observable at the $B$ factories.  The channels involving
right and wrong charm neutral vector mesons turn out to be most
promising: they are expected to have the largest asymmetries, and the
theoretical method yields the best results for the production rates
and spectra of the vector mesons~\cite{CMS99}.

\section*{Acknowledgments}
The authors thank Thomas Gehrmann for fruitful discussions.  This work
(X.~C. during his time in Karlsruhe, T.~M. and I.~S.) was supported by
the DFG Graduiertenkolleg ``Elementarteilchenphysik an
Beschleunigern'' and by the DFG Forschergruppe ``Quantenfeldtheorie,
Computeralgebra und Monte-Carlo-Simulation.''


\end{document}